\documentclass{article}

\usepackage{microtype}
\usepackage{graphicx}
\usepackage{subcaption}
\usepackage{booktabs} 

\usepackage{hyperref}

\usepackage[preprint]{icml2026}

\usepackage{amsmath}
\usepackage{amssymb}
\usepackage{mathtools}
\usepackage{amsthm}
\usepackage{booktabs}
\usepackage{multirow}
\usepackage{graphicx}
\usepackage[table]{xcolor}
\usepackage{booktabs}
\usepackage{multirow}
\usepackage{graphicx}
\definecolor{bestred}{HTML}{C00000}
\definecolor{secondblue}{HTML}{0070C0}

\newcommand{\best}[1]{\textcolor{bestred}{\textbf{#1}}}
\newcommand{\second}[1]{\textcolor{secondblue}{\textbf{#1}}}
\usepackage[capitalize,noabbrev]{cleveref}

\theoremstyle{plain}

\theoremstyle{definition}

\theoremstyle{remark}

\usepackage[textsize=tiny]{todonotes}

\icmltitlerunning{Vision–Language Controlled Deep Unfolding for Joint Medical Image Restoration and Segmentation}

\begin{document}

\twocolumn[

\icmltitle{Vision–Language Controlled Deep Unfolding for Joint Medical Image Restoration and Segmentation}

  \icmlsetsymbol{equal}{*}

  \begin{icmlauthorlist}
    \icmlauthor{Ping Chen}{equal,yyy}
    \icmlauthor{Zicheng Huang}{equal,sch}
    \icmlauthor{Xiangming Wang}{equal,yyy}
    \icmlauthor{Yungeng Liu}{yyy}
    \icmlauthor{Bingyu Liang}{yyy}
    \icmlauthor{Haijin Zeng}{comp}
    \icmlauthor{Yongyong Chen}{yyy}
  \end{icmlauthorlist}

  \icmlaffiliation{yyy}{Harbin Institute of Technology (Shenzhen)}
  \icmlaffiliation{comp}{Harvard University}
  \icmlaffiliation{sch}{SUN YAT-SEN UNIVERSITY}

  \icmlcorrespondingauthor{Haijin Zeng}{haijin.zeng@imec.be}
  \icmlcorrespondingauthor{Yongyong Chen}{YongyongChen.cn@gmail.com}

  \icmlkeywords{AiOMIRS, Unfolding, Vision-Language}

  \vskip 0.3in
]



\printAffiliationsAndNotice{}  

\begin{abstract}
We propose VL-DUN, a principled framework for joint All-in-One Medical Image Restoration and Segmentation (AiOMIRS) that bridges the gap between low-level signal recovery and high-level semantic understanding. While standard pipelines treat these tasks in isolation, our core insight is that they are fundamentally synergistic: restoration provides clean anatomical structures to improve segmentation, while semantic priors regularize the restoration process. VL-DUN resolves the sub-optimality of sequential processing through two primary innovations. (1) We formulate AiOMIRS as a unified optimization problem, deriving an interpretable joint unfolding mechanism where restoration and segmentation are mathematically coupled for mutual refinement. (2) We introduce a frequency-aware Mamba mechanism to capture long-range dependencies for global segmentation while preserving the high-frequency textures necessary for restoration. This allows for efficient global context modeling with linear complexity, effectively mitigating the spectral bias of standard architectures.
As a pioneering work in the AiOMIRS task, VL-DUN establishes a new state-of-the-art across multi-modal benchmarks, improving PSNR by 0.92 dB and the Dice coefficient by 9.76\%. Our results demonstrate that joint collaborative learning offers a superior, more robust solution for complex clinical workflows compared to isolated task processing. The codes are provided in \href{https://github.com/cipi666/VLDUN}{GitHub}.
\end{abstract}

\section{Introduction}

Medical image analysis plays a pivotal role in modern clinical workflows, primarily encompassing tasks such as Medical Image Restoration (MedIR) and Medical Image Segmentation (MedIS). Recently, the concept of All-in-One Medical Image Task (AiOMIT) has gained traction, aiming to address diverse problems across multiple modalities within a unified framework. While significant progress has been made in All-in-One MedIR (AiOMedIR)~\cite{yang2024all, yang2025tat} and All-in-One MedIS (AiOMedIS)~\cite{ma2024segment, SAM2}, bridging these two domains to achieve \textbf{All-in-One Medical Image Restoration and Segmentation (AiOMIRS)} remains a formidable challenge.

\begin{figure}[htbp]
  \centering
  \includegraphics[width=0.48\textwidth]{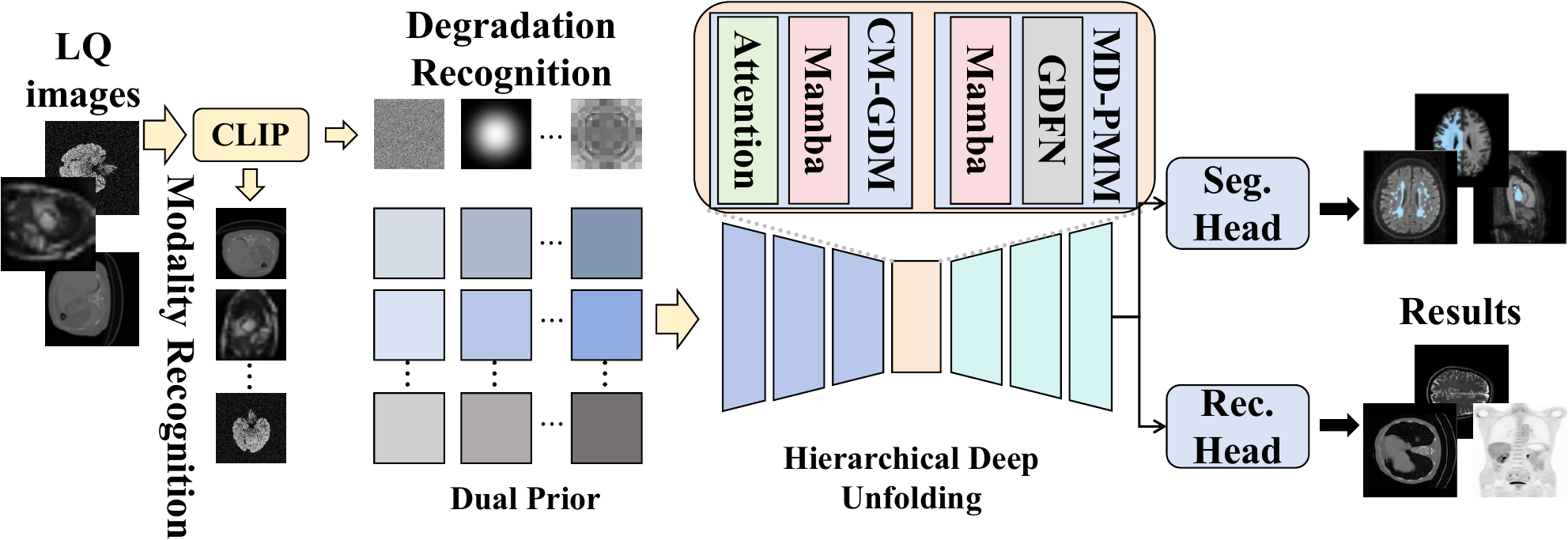} 
  \caption{Our VL-DUN leverages CLIP to extract explicit dual priors regarding modality and degradation. Our DUN uses frequency-decoupling strategy, the low-frequency feature is processed by GDFN, while Mamba captures high-frequency texture.}
  \label{fig:overall_motivation}
\end{figure}

AiOMedIR aims to reconstruct high-quality (HQ) images $x$ from low-quality (LQ) observations $y$ such as MRI, CT, X-Ray. The imaging process is typically modelled as an ill-posed inverse problem:
\begin{equation}
\label{eqn-1}
    y = \Phi{x}+n, \quad n \sim \mathcal{N}(0, \Sigma),
\end{equation}
where $\Phi$ denotes the degradation matrix and $n$ represents noise. Concurrently, MedIS aims to assign a semantic label $s_{i,j}$ to each pixel $(i,j)$ of the image $x$, mapping the anatomical structure to a segmentation mask $S$. However, in real-world clinical scenarios, images are often degraded ($y$). Standard segmentation models~\cite{DBLP:conf/miccai/BozorgpourKAHM25, DBLP:conf/miccai/ChenQCHFWL25} trained on HQ data perform poorly on LQ inputs due to domain shift and artifact interference. Therefore, AiOMIRS requires a model to simultaneously solve Eq.~\eqref{eqn-1} and predict $S$ from $y$.

Existing AiO methods~\cite{yang2024all,yang2025tat,ma2024segment} treat image restoration and image segmentation as two separate task, ignoring their potential connections. Although some works~\cite{DBLP:conf/eccv/BuchholzPSKJ20, DBLP:conf/miccai/WangLCSDHX24, DBLP:conf/miccai/YeNHWLLSHWL25} have studied the tasks of reconstruction and segmentation simultaneously, they have not addressed the problem from a multi-modal perspective. First, pure deep learning-based methods~\cite{yang2024all, potlapalli2023promptir, li2022airnet} operate as black boxes, lacking the interpretability essential for medical diagnosis. While Deep Unfolding Networks (DUNs)~\cite{zhang2017learning, zhang2018ista, mou2022deep} offer interpretability by unrolling optimization algorithms, they typically struggle with the severe \textbf{distribution shifts} present in multi-modality data.

To address these challenges, we propose the \textbf{Vision-Language Controlled Deep Unfolding Network (VL-DUN)}. Unlike previous methods that treat tasks in isolation, we fundamentally reformulate AiOMIRS as a unified optimization problem. Specifically, to tackle the severe \textbf{distribution shifts} inherent in multi-modal data, we introduce a \textbf{Vision-Language Dual Prior} mechanism. By leveraging a fine-tuned CLIP model~\cite{radford2021learning} trained on 8 diverse medical datasets, we extract explicit \textit{Modality} and \textit{Degradation Priors}. These priors are dynamically injected into the unfolding stages to calibrate the degradation operator, enabling the network to cognitively adapt to varying clinical distributions and artifact patterns.

Furthermore, to implement this unified framework with efficient global context modeling, we design a hierarchical architecture based on \textbf{Frequency-Aware Mamba}. While Mamba~\cite{gu2023mamba, DBLP:conf/cvpr/ErgastiBFFB025} offers linear complexity suitable for processing high-resolution medical images, standard architectures suffer from spectral bias, often losing the high-frequency textures crucial for restoration. 
To reconcile this, our Frequency-Aware mechanism decouples the feature space: it leverages Mamba to capture long-range dependencies for accurate global segmentation, while explicitly preserving high-frequency components for high-fidelity restoration. This synergistic design overcomes the limitations of traditional CNNs lacking of global context~\cite{DBLP:conf/miccai/LinYDZY23} and Transformers with quadratic complexity, offering a robust solution for the AiOMIRS task.

Our main contributions are summarized as follows:
\begin{itemize}
    \item We define and explore the challenging AiOMIRS task, and propose the Vision-Language Controlled Deep Unfolding Network (VL-DUN). 
    
    \item We leverage a fine-tuned CLIP on 8 different medical datasets to extract explicit \textit{Modality} and \textit{Degradation Priors}. Our CLIP has strong ability to recognize modality and degradation features.
    
    \item We design a novel \textbf{Frequency-Aware Mamba Proximal Module}. By decoupling high-frequency texture, this design overcomes spectral bias and achieves global context modeling with linear complexity.

    \item Extensive experiments demonstrate that our model achieves state-of-the-art in AiOMIRS task by improving 0.92 in PSNR and 9.76\% in Dice. We also theoretically prove the benefits of joint tasks in Appendix~\ref{sec:synergy_analysis}.
\end{itemize}

\section{Related Work}
\label{sec:related_work}

\subsection{All-in-One Medical Image Task}
Medical image analysis has traditionally treated restoration and segmentation as independent tasks, optimizing them separately for specific modalities. Recently, the concept of \textbf{AiOMIT} has emerged, aiming to develop unified models capable of handling diverse modalities and tasks simultaneously.
In the realm of restoration, all-in-one methods like AirNet~\cite{li2022all} and recent approaches such as AMIR~\cite{yang2024all} and TAT~\cite{yang2025tat} attempt to remove various degradations across different conditions using a unified network. Similarly, generic segmentation models such as nnU-Net~\cite{isensee2021nnu} and recent foundation models like SAM series~\cite{kirillov2023segment, ravi2024sam2,carion2025sam3segmentconcepts} and MedSAM series~\cite{ma2024segment, SAM2,liu2025medsam3} have demonstrated impressive zero-shot segmentation capabilities across different organs and modalities. 

Despite these advancements, existing AiO frameworks typically focus on either restoration or segmentation in isolation. In real-world clinical scenarios, however, low-quality (LQ) inputs significantly degrade the performance of downstream segmentation. Current joint restoration-and-segmentation methods are mostly limited to single-modality tasks and they just focus on single degradation type~\cite{DBLP:conf/eccv/BuchholzPSKJ20, DBLP:conf/miccai/WangLCSDHX24}. \textbf{In contrast, our work bridges this gap by proposing a unified AiOMIRS framework that simultaneously performs high-fidelity restoration and precise segmentation across multiple modalities, leveraging restoration gains to boost downstream tasks.}

\subsection{Vision-Language Models for MedIR}
The success of \textbf{Vision-Language Models (VLMs)}, particularly CLIP~\cite{radford2021learning}, has inspired a new wave of research in low-level vision. Text prompts have been utilized to guide image restoration networks, providing semantic cues about degradation types. Methods like PromptIR~\cite{potlapalli2023promptir} and DA-CLIP~\cite{luo2024controlling} employ learnable prompts or controllers to adapt models to different degradations in natural images.

However, direct application of natural image VLMs to medical restoration faces two challenges: substantial domain gaps and the lack of fine-grained degradation awareness in medical contexts. Most existing methods rely on coarse prompts or fixed embeddings. Although  recent works try to fine-tune VLMs in medical domain~\cite{jiang2025hulumedtransparentgeneralistmodel, DBLP:journals/corr/abs-2510-15418}, they are not suitable for low-level vision tasks. \textbf{Differently, we fine-tune CLIP specifically on 8 medical datasets to construct a Dual Prior mechanism. We explicitly extract both \textit{Modality Priors} and \textit{Degradation Priors}, controlling restoration process.}

\subsection{Deep Unfolding Networks}
\textbf{Deep Unfolding Networks (DUNs)}~\cite{zhang2017learning, zhang2018ista} have gained popularity for their unique ability to combine model-based optimization with the learning power of neural networks. By unrolling iterative algorithms such as Alternating Direction Method of Multipliers (ADMM), Proximal Gradient Descent (PGD) into deep architectures, DUNs provide transparent intermediate results and controllable restoration processes.

Existing DUNs typically employ CNNs or Transformers~\cite{mou2022deep, DBLP:conf/aaai/WangZC0CC25} as the proximal operator denoiser. While CNNs are computationally efficient, they struggle to capture long-range dependencies crucial for restoring global anatomical structures. Transformers offer global context but suffer from quadratic computational complexity $\mathcal{O}(N^2)$, making them computationally prohibitive for high-resolution medical images. \textbf{To address this trade-off, we propose the  Attention-Mamba based Deep Unfolding Network. By integrating the Mamba~\cite{gu2023mamba} into the unfolding framework, we achieve linear complexity $\mathcal{O}(N)$ with global receptive fields.} Recently, VLU-Net~\cite{vlunet2025} uses CLIP to guide their DUN as well. However, their CLIP only extracts latent degradation prior in nature image, which is short in medical images due to different modality features and data distributions. Therefore, we design a novel mechanism that dynamically modulates the optimization based on our Vision-Language dual Priors, ensuring adaptability to unknown degradation and different data distributions.

\section{Method}

\begin{figure*}[htbp]
  \centering
  \includegraphics[width=1\textwidth]{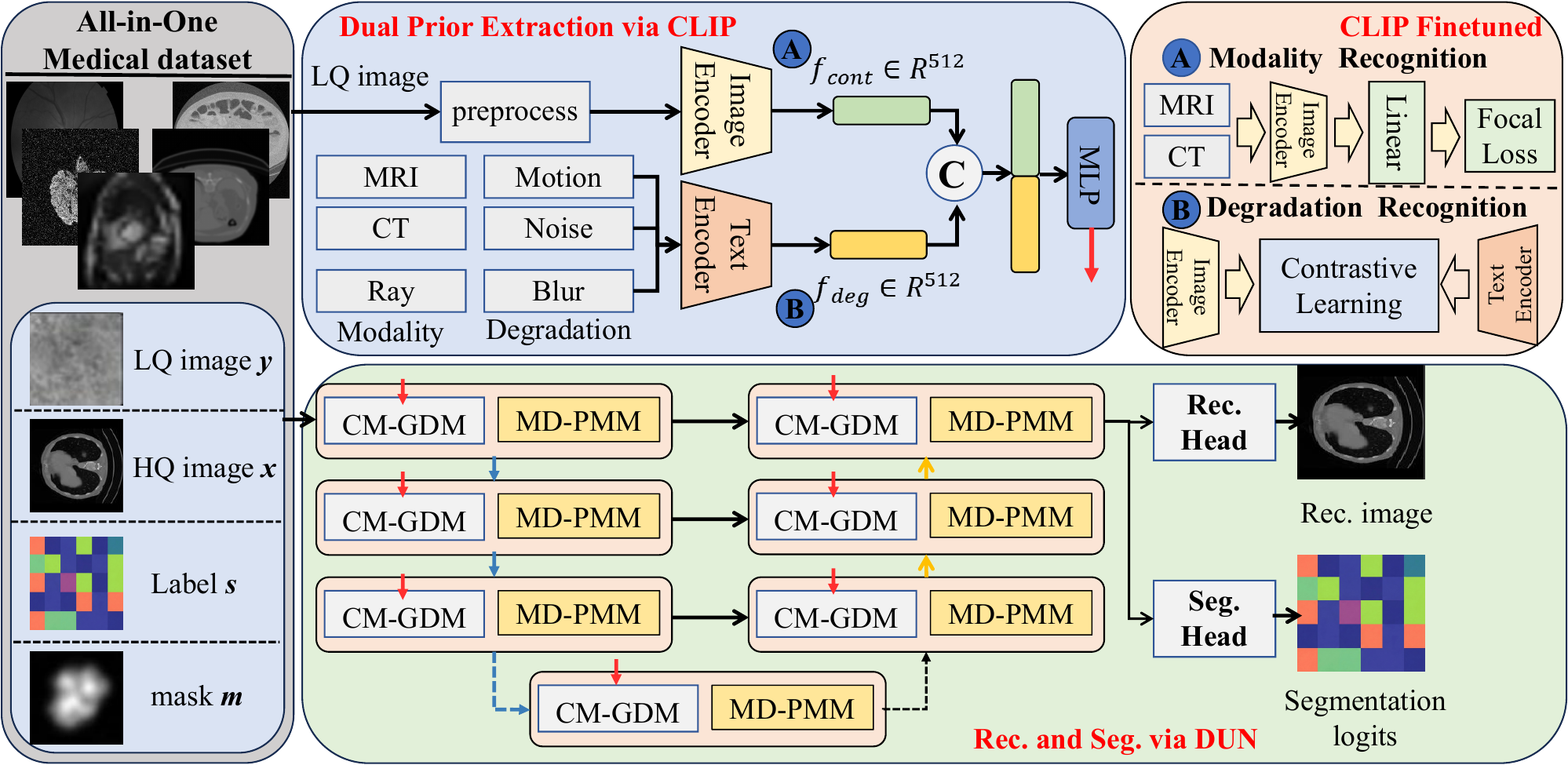} 
  \caption{\textbf{The overall architecture of the proposed VL-DUN.} Our VL-DUN comprises two main components: (1) \textbf{The Vision-Language Prior Extraction Module (Top):} Given a LQ input, it utilizes a fine-tuned CLIP model to extract explicit semantic cues. (2) \textbf{The Attention-Mamba based Deep Unfolding Network (Bottom):} Mathematically grounded in the Proximal Gradient Descent algorithm, the network unfolds the optimization into $K$ stages arranged in a hierarchical architecture.}
  \label{fig:overview}
\end{figure*}

\subsection{Problem Definition}

The goal of AiOMIRS is to simultaneously recover a HQ image $x \in \mathbb{R}^{H \times W \times C}$ and predict a semantic segmentation mask $S \in \{0, 1\}^{H \times W \times K}$ from a given LQ image $y$. The degradation process can be formulated as a linear inverse problem as Eq.~\eqref{eqn-1}. Our objective is to leverage prior knowledge to explicitly identify the latent attributes of $\Phi$ and use them to guide the restoration and segmentation process.

\subsection{Overview of our Architecture}

As illustrated in Fig.~\ref{fig:overview}, our proposed \textbf{Vision-Language Controlled Deep Unfolding Network (VL-DUN)} constitutes a unified framework designed for the challenging AiOMIRS task. The architecture integrates semantic understanding from CLIP with the interpretability of model-based optimization. Our pipeline consists of two components:

\textbf{Vision-Language Prior Extraction.} Given a LQ input $y$, we first utilize a fine-tuned CLIP model to extract \textit{Modality Prior} and \textit{Degradation Prior}.
    
\textbf{Attention-Mamba based Deep Unfolding.} The restoration network unfolds the iterative optimization process into $K$ stages arranged in a hierarchical architecture. Each stage $k$ updates $x_k$ through two modules. The Attention-Mamba union Gradient Descent Module (AMGDM), which dynamically estimates the degradation operator $\Phi$ and updates the data fidelity term using the injected semantic priors; and the Mamba-GDFN based Proximal Map Module (MDPMM), which projects the intermediate result onto the clean image manifold, utilizing a frequency-decoupled Mamba strategy for efficient and structure-preserving denoising. At the end of the network, we branch into the Rec. Head and Seg. Head to complete AiOMIRS tasks.

\subsection{Vision-Language Prior Extraction}

\subsubsection{Motivation}

Existing deep learning-based MedIR models process inputs indiscriminately, relying on a static set of parameters to handle diverse distributions and degradation patterns. To address this, we propose a Vision-Language Dual Prior mechanism. We argue that visual perception and semantic understanding via text description are complementary. By injecting these priors, we transform the restoration problem into a guided process. We leverage CLIP as our backbone due to its robust representation capabilities, and we fine-tune it specifically for medical scenarios. More analysis that leverage CLIP are demonstrated in Appendix~\ref{sec:CLIP}.

\begin{figure}[htbp]
  \centering
  \includegraphics[width=0.47\textwidth]{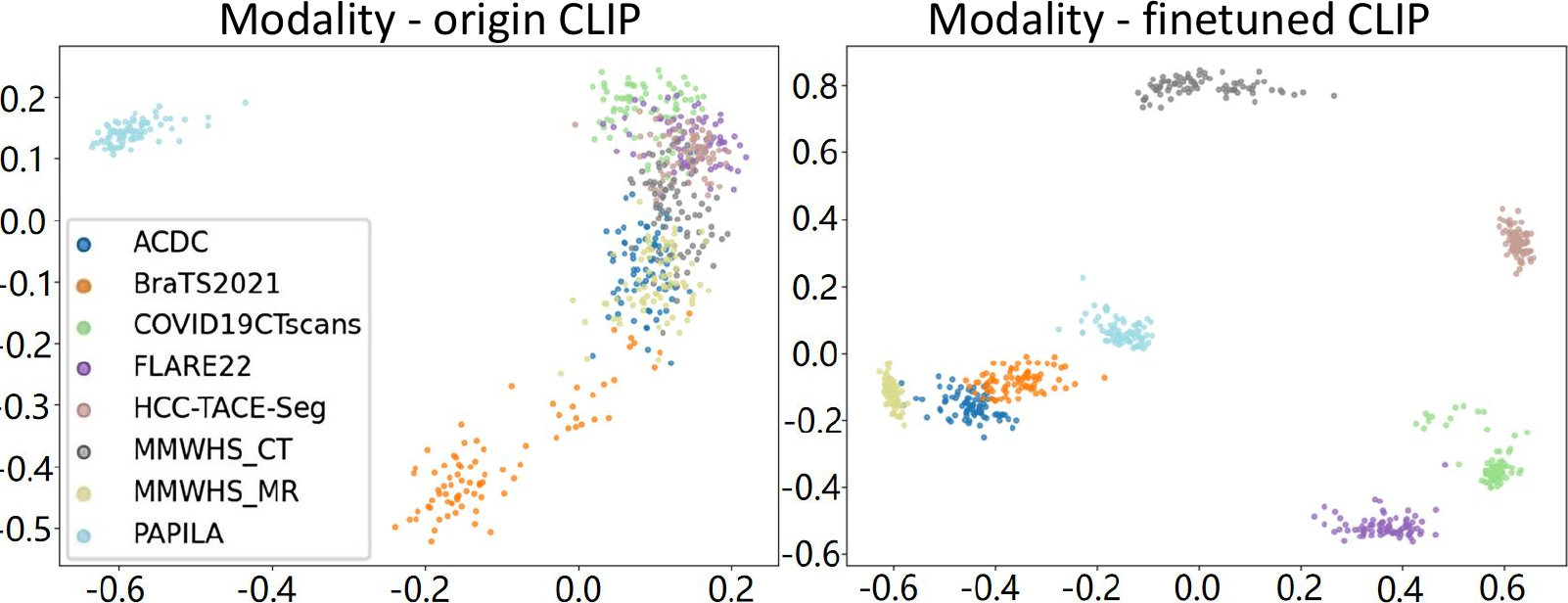} 
  \caption{By comparison, our fine-tune CLIP can distinguish 8 modalities much better than basic CLIP.}
  \label{fig:Modality_Finetune}
\end{figure}

\subsubsection{Visual Prior for Modality Recognition}

Medical images from different modalities exhibit severe distribution shifts. To inform the network about the data distribution it is processing, we construct a Visual Prior branch. We treat modality recognition as a supervised classification task. Given a set of $M$ diverse medical datasets, each dataset represents a distinct modality class. We utilize the CLIP visual encoder $E_v$ followed by a learnable linear classifier $W_{cls}$. Since medical datasets are often extremely imbalanced in size, we employ Focal Loss to optimize this branch. Formally, given an input image $y$, the modality feature $f_{mod}$ and the predicted probability $p$ is computed as:
\begin{equation}
f_{mod} = E_v(y), \quad p = \text{softmax}(W_{cls} f_{mod}).
\end{equation}

The optimization objective $\mathcal{L}_{mod}$ is defined as:
\begin{equation}
\mathcal{L}_{mod} = -\alpha (1 - p_t)^\gamma \log(p_t),
\end{equation}
where $p_t$ is the probability of the true class, and $\alpha, \gamma$ are focusing parameters. The resulting modality vector serves as a modality prior. As shown in Fig.~\ref{fig:Modality_Finetune}, we fine-tune our CLIP across 8 different modalities, and the comparison between the original and our fine-tuned CLIP demonstrates that our fine-tuned CLIP can recognize different modalities.

\subsubsection{Text Prior for Degradation Recognition}

\begin{figure}[htbp]
  \centering
  \includegraphics[width=0.48\textwidth]{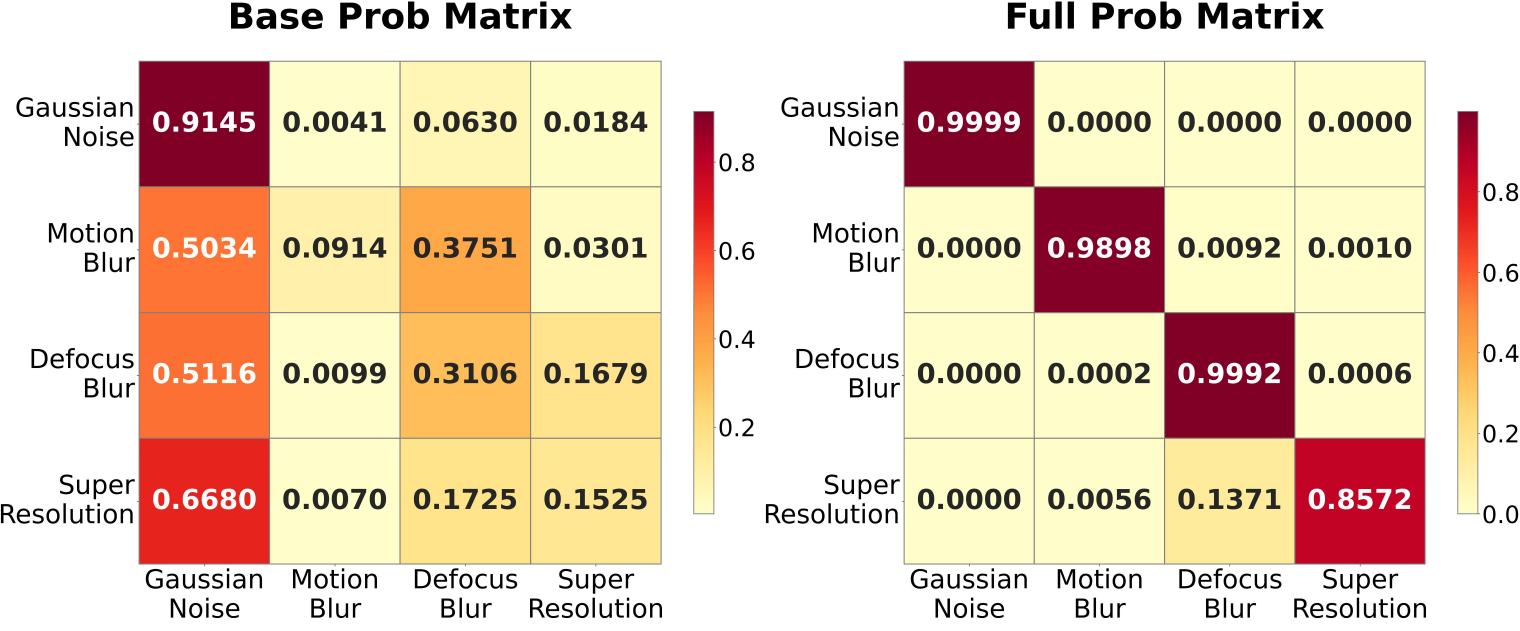} 
  \caption{Although the original CLIP has high performance in natural image degradation recognition, it lacks generality in the medical image degradation recognition scene. Our fine-tuned CLIP is able to recognize different degradation well.}
  \label{fig:Degradation_Finetune}
\end{figure}

\begin{figure*}[htbp]
  \centering
  \includegraphics[width=1\textwidth]{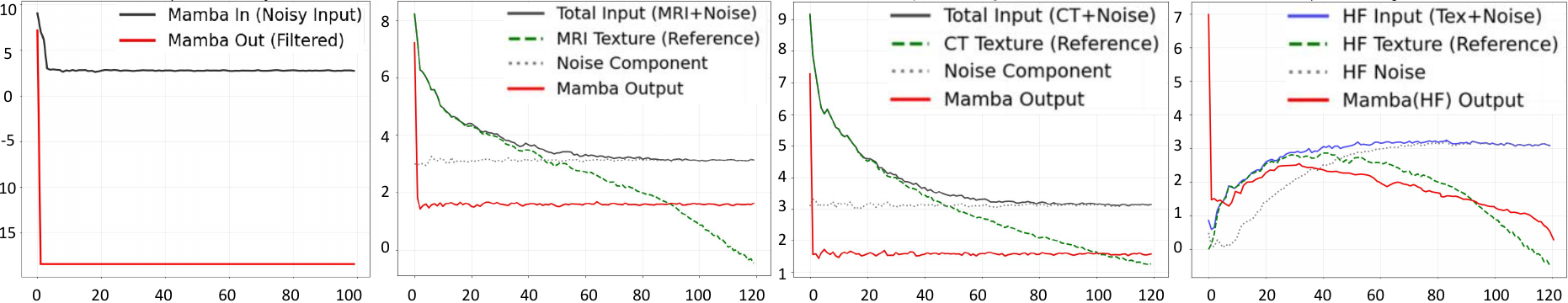} 
  \caption{Spectral analysis reveals that standard Mamba exhibits inherent low-pass filtering properties, suppressing high-frequency textures alongside noise. In contrast, our frequency-decoupled strategy enables it to effectively distinguish anatomical details from noise by leveraging global context on the high-frequency band.}
  \label{fig:mamba}
\end{figure*}

While modality features capture global distribution, they lack specific information about artifacts. To this end, we introduce a Text Prior branch to identify the degradation type such as noise, blur, motion, low-resolution. Unlike the modality branch, which relies on pure visual classification, here we leverage the vision-language alignment capability of CLIP. We define a set of $K$ fixed textual prompts describing specific degradations such as ``Noisy medical image with severe grain''. These prompts serve as semantic anchors. During training, we synthesize degraded images on-the-fly to create pairs of degraded image $y$ with degradation label $k$. We freeze the CLIP text encoder $E_t$ to preserve its pre-trained semantic space and fine-tune the visual encoder $E_v$. The similarity score $s_{i,k}$ between the image feature and the $k$-th text prompt feature is calculated via dot product. We employ an Image-Text Contrastive Loss implemented as Cross-Entropy over similarity scores to align the visual artifacts with their textual descriptions.

The resulting similarity distribution over the text prompts provides a semantically rich degradation prior. As shown in Fig.~\ref{fig:Degradation_Finetune}, we fine-tune our CLIP across four different degradation types, and the comparison between origin and our fine-tuned CLIP demonstrates that our fine-tuned CLIP has ability to recognize different degradation types.

\subsection{Attention-Mamba based Deep Unfolding Network}

\begin{figure}[htbp]
  \centering
  \includegraphics[width=0.45\textwidth]{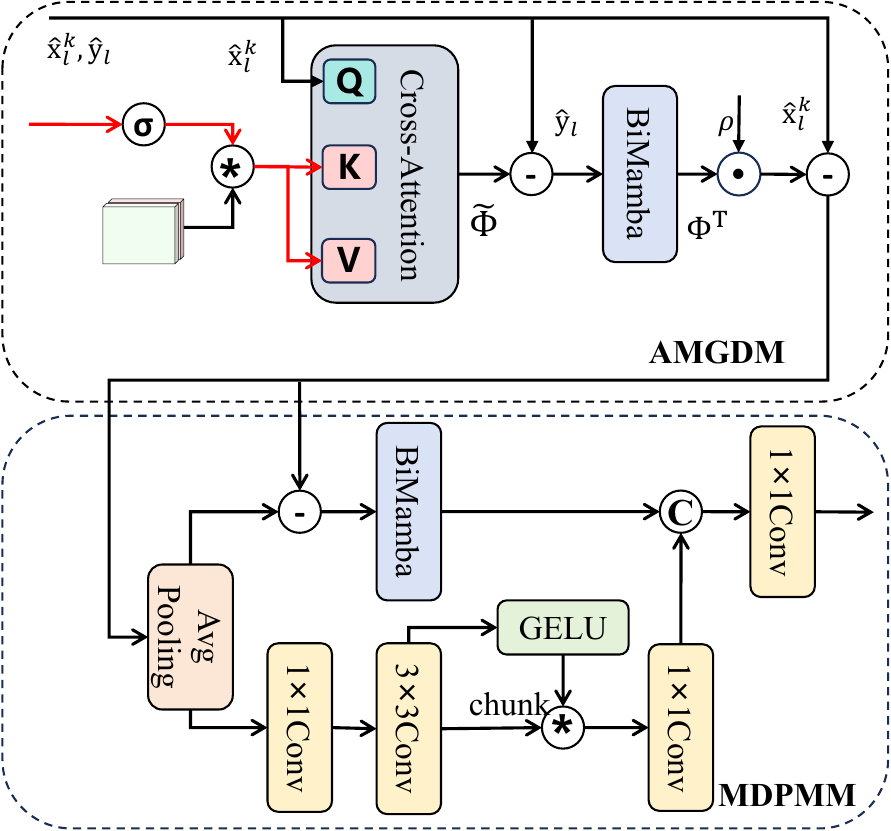} 
  \caption{\textbf{The detailed components of AMGDM and MDPMM.} Our AMGDM leverage the advantages of Cross-Attention and BiMamba, and our MDPMM further completes denoising via frequency-decoupling strategy.}
  \label{fig:o}
\end{figure}

\subsubsection{Overview and Unfolding Formulation}
To reconcile the trade-off between global modeling and computational efficiency, we integrate the linear-complexity Mamba into DUN. Following the Proximal Gradient Descent (PGD) algorithm, we minimize the energy function $\hat{x} = \arg\min_x \frac{1}{2} \|y - \Phi x\|^2 + \lambda \mathcal{R}(x)$. The optimization is unfolded into $K$ stages within a hierarchical architecture. The detailed architecture is shown in Fig.~\ref{fig:o}.

\subsubsection{Dual Prior Guided AMGDM}
The AMGDM dynamically updates the data fidelity term. To address the unknown degradation $\Phi$, we leverage the extracted Vision-Language Priors. First, we utilize a Cross-Attention mechanism where image features serve as Queries and the CLIP-projected prior serves as Keys and Values. This allows the network to simulate the specific degradation pattern, described by the text prompt. Additionally, to map the global residual error $\Phi x - y$, we employ a \textbf{Bi-directional Mamba Layer}. This enables precise gradient estimation by aggregating global error information.

\subsubsection{Frequency-Aware MDPMM}
The Proximal Module aims to project the estimation onto the clean image manifold. As shown in Fig.~\ref{fig:mamba}, standard Mamba exhibits inherent spectral bias, acting as a low-pass filter that tends to over-smooth high-frequency textures. Direct application fails to distinguish between noise and textures. To address this, we propose a Frequency-Aware Mamba Block that decouples the restoration process.
\begin{equation}
    x_{low} = \text{AvgPool}(z_k), \quad x_{high} = z_k - x_{low}.
\end{equation}
\textbf{High-Frequency Branch.} We handle $x_{high}$ into a Bi-directional Mamba layer. By isolating high frequencies, we force Mamba to utilize its global receptive field to differentiate between spatially correlated anatomical textures and independent noise patterns, preventing texture loss.

\textbf{Low-Frequency Branch.} Since low frequencies represent stable anatomical structures, we process $x_{low}$ with a lightweight Gated-Dconv Feed-Forward Network (GDFN) to maintain structural consistency.

\begin{figure*}[t]
  \centering
  \includegraphics[width=1\textwidth]{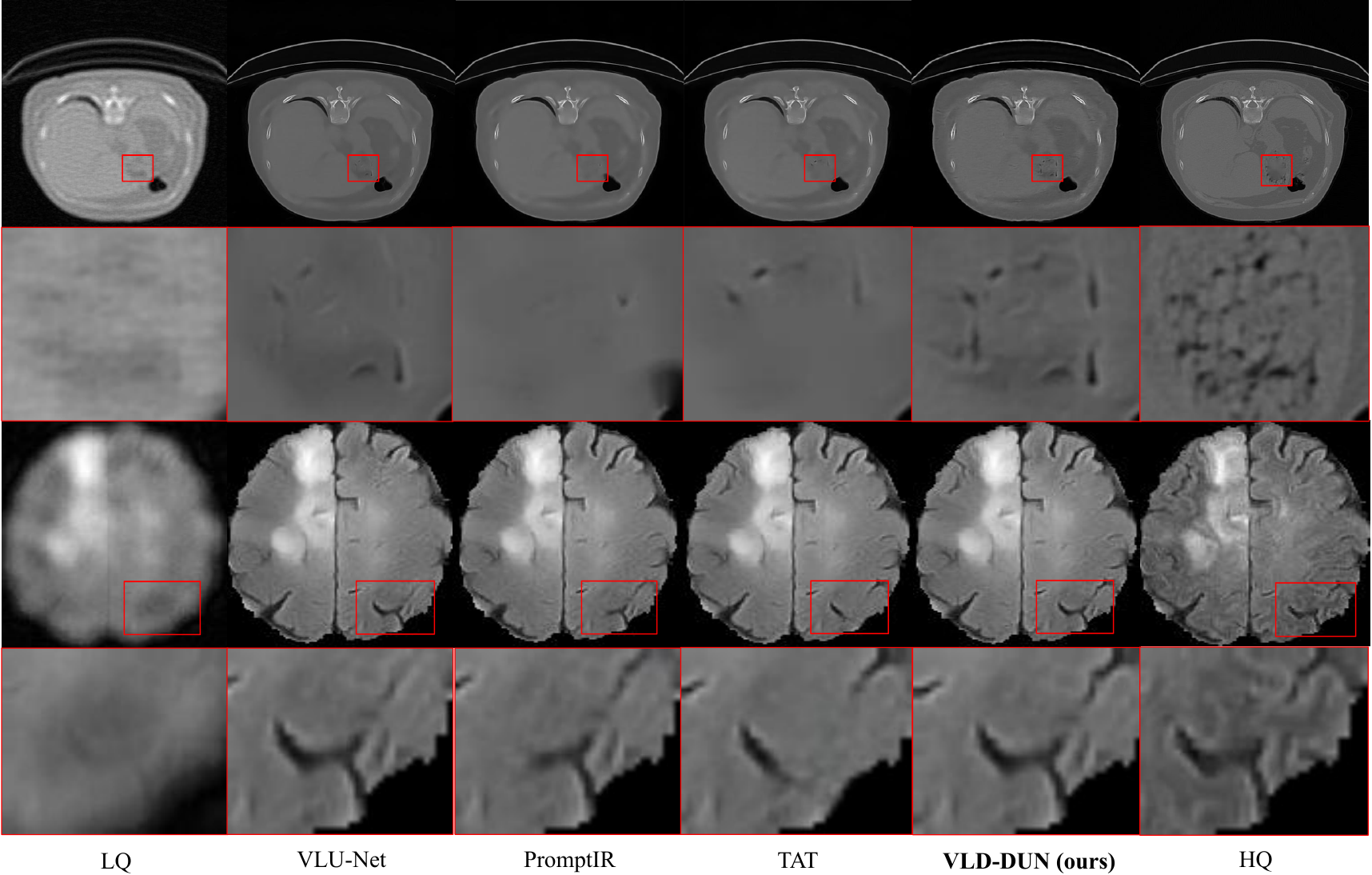} 
  \caption{\textbf{Visualization of Comparative Experiments.}  We provided visualizations of BrainTS2021 and COVID19-CT, and entire experiments are conducted on RTX pro 6000. Our VL-DUN is able to reconstruct more details and achieves state-of-the-art performance in reconstruction.}
  \label{fig:visual}
\end{figure*}

Finally, the dual branches are fused via a convolution. This design overcomes spectral bias while reducing computational redundancy. To achieve joint optimization, we attach a lightweight segmentation head to the final unfolding stage. This design is feasible because the unfolding process recovers high-fidelity structures, providing rich features that benefit the segmentation task without requiring a separate encoder. The further analysis is provided in Appendix~\ref{sec:dun_ab}.

\section{Experiment}

\begin{table*}[t]
    \centering
    \caption{\textbf{Quantitative comparison on four medical imaging datasets.} We compare our VL-DUN with state-of-the-art restoration and joint-task methods. \best{Red} indicates the best performance, and \second{Blue} indicates the second-best performance. ``-'' denotes that the method does not support the corresponding metric. The row with a gray background highlights our proposed method.}
    \label{tab:quant_comparison}
    \resizebox{\textwidth}{!}{
        \begin{tabular}{c|ccc|ccc|ccc|ccc|ccc}
            \toprule
            \multirow{2}{*}{\textbf{Method}} & \multicolumn{3}{c|}{\textbf{ACDC}} & \multicolumn{3}{c|}{\textbf{COVID19CTscans}} & \multicolumn{3}{c|}{\textbf{HCC-TACE-Seg}} & \multicolumn{3}{c|}{\textbf{BraTS2021}} & \multicolumn{3}{c}{\textbf{Average}} \\
            \cmidrule(lr){2-4} \cmidrule(lr){5-7} \cmidrule(lr){8-10} \cmidrule(lr){11-13} \cmidrule(lr){14-16}
             & PSNR$\uparrow$ & SSIM$\uparrow$ & Dice$\uparrow$ & PSNR$\uparrow$ & SSIM$\uparrow$ & Dice$\uparrow$ & PSNR$\uparrow$ & SSIM$\uparrow$ & Dice$\uparrow$ & PSNR$\uparrow$ & SSIM$\uparrow$ & Dice$\uparrow$ & PSNR$\uparrow$ & SSIM$\uparrow$ & Dice$\uparrow$ \\
            \midrule
            DenoiSeg & 22.35 & 0.5943 & \second{0.6423} & 12.96 & 0.3984 & 0.5929 & 24.25 & 0.7050 & \second{0.6001} & 26.23 & 0.5921 & \second{0.4132} & 21.45 & 0.5725 & \second{0.5621} \\
            PromptIR & 24.40 & 0.6772 & - & 21.95 & 0.5138 & - & 33.14 & 0.9108 & - & 31.27 & 0.8596 & - & 27.69 & 0.7404 & - \\
            EM-DenoiSeg & 22.07 & 0.5395 & 0.6215 & 16.42 & 0.3862 & 0.5947 & 25.86 & 0.7154 & 0.5951 & 27.10 & 0.6739 & {0.3945} & 22.86 & 0.5788 & 0.5515 \\
            AMIR & 24.32 & 0.6753 & - & 21.45 & 0.5111 & - & 33.14 & 0.9112 & - & 31.32 & 0.8601 & - & 27.56 & 0.7394 & - \\
            VLU-Net & \second{25.45} & \second{0.7098} & - & 20.65 & \second{0.5228} & - & \second{33.84} & \best{0.9344} & - & \best{32.65} & \best{0.8783} & - & \second{28.15} & \best{0.7613} & - \\
            TAT & 24.94 & 0.7066 & - & \second{22.12} & 0.5124 & - & 33.16 & 0.9201 & - & 31.14 & 0.8596 & - & 27.84 & 0.7497 & - \\
            
            \rowcolor{gray!20} \textbf{VL-DUN (Ours)} & \best{25.93} & \best{0.7129} & \best{0.7563} & \best{22.40} & \best{0.5330} & \best{0.7742} & \best{35.63} & \second{0.9239} & \best{0.6401} & \second{32.31} & \second{0.8625} & \best{0.4680} & \best{29.07} & \second{0.7581} & \best{0.6597} \\
            \bottomrule
        \end{tabular}
    }
\end{table*}

\subsection{Datasets and Benchmarks}
\label{sec:datasets}

To validate the effectiveness of our proposed VL-DUN, we construct a comprehensive benchmark covering diverse medical modalities and anatomical structures. The data usage is divided into two phases. CLIP Pre-training and Joint Task Evaluation. Our datasets and benchmark are based on IMIS-bench~\cite{cheng2024interactivemedicalimagesegmentation}.

\textbf{Datasets for CLIP Fine-tuning.} 
To equip our Vision-Language Prior module with robust distribution awareness, we fine-tune the CLIP visual encoder on a large-scale collection of \textbf{8 medical datasets}. These include ACDC~\cite{bernard2018deep}, BraTS2021~\cite{baid2021rsna}, and MMWHS-MR~\cite{zhuang2016multi}, COVID19-CT~\cite{ma2021towards}, FLARE22~\cite{ma2024fast}, HCC-TACE-Seg~\cite{ma2024segment}, and MMWHS-CT~\cite{zhuang2016multi} (Whole Heart), PAPILA~\cite{kovacs2020papila}.
This diverse collection ensures that the extracted Modality Prior is generic and discriminative across major medical images.

\textbf{Datasets for Joint Restoration and Segmentation.} 
For the main AiOMIRS task, we select a representative subset of \textbf{4 datasets} to evaluate the model's performance on core clinical tasks. \textbf{ACDC~\cite{bernard2018deep},} Containing 100 patients with cine-MRI, we perform 4-class segmentation (RV, LV, Myocardium, Background). \textbf{BraTS2021~\cite{baid2021rsna},} A large-scale brain tumor dataset, we utilize the T1 modality for restoration and segment the tumor regions. \textbf{COVID19-CT~\cite{ma2021towards},} consisting of lung CT scans with infection masks, We perform binary segmentation of the infection lesions. \textbf{HCC-TACE-Seg~\cite{moawad2021hcc},} focusing on liver tumor segmentation in contrast-enhanced CT scans.

\textbf{Degradation Simulation.} 
To unify the physical operator formulation across different modalities and reduce computational overhead, we model the degradation for both CT and MRI as an undersampling process in the frequency domain. Let $\mathbf{x}$ denote the ground truth and $\mathbf{y}$ be the LQ observation. The degradation process is formulated as:
\begin{equation}
    \mathbf{y} = \mathcal{F}^{-1} ( \mathbf{M} \odot \mathcal{F}(\mathbf{x}) ),
    \label{eq:freq_degradation}
\end{equation}
where $\mathcal{F}$ denotes the Fast Fourier Transform (FFT), $\mathcal{F}^{-1}$ is its inverse, and $\odot$ represents element-wise multiplication. $\mathbf{M}$ is a binary mask determining the sampling trajectory, which varies by modality. For CT, based on the Central Slice Theorem~\cite{kak1988principles}, data acquired from parallel-beam projections at specific angles corresponds to radial lines passing through the origin in the 2D frequency domain. Therefore, we simulate sparse-view CT artifacts by defining $\mathbf{M}$ as a radial mask. This approach allows us to emulate the physics of sparse-view reconstruction directly in the frequency domain. For MRI, we apply Cartesian undersampling masks for $\mathbf{M}$ in the k-space~\cite{lustig2007sparse}. This generates aliased inputs that mimic realistic accelerated MRI acquisition scenarios.

\subsection{AiOMIRS task}

In this section, we conduct comparative experiments on AiOMedIRS, with the model input being LQ images. Our baseline models include DenoiSeg~\cite{DBLP:conf/eccv/BuchholzPSKJ20}, PromptIR~\cite{potlapalli2023promptir}, AMIR~\cite{yang2024all}, EM-DenoiSeg~\cite{DBLP:conf/miccai/WangLCSDHX24}, VLU-Net~\cite{vlunet2025}, TAT~\cite{yang2025tat}.

\begin{figure}[t]
  \centering
  \includegraphics[width=0.48\textwidth]{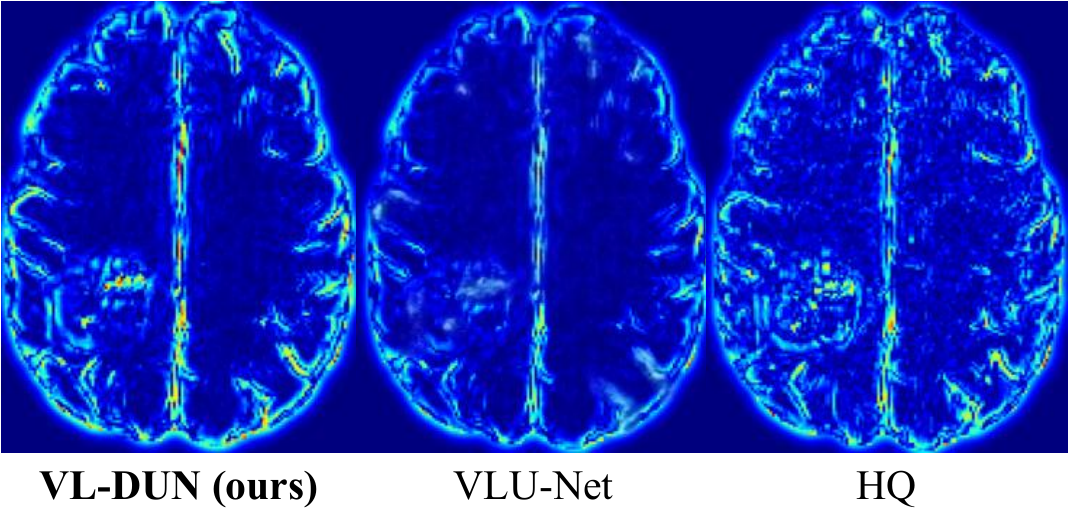} 
  \caption{Visualization of our VL-DUN in high-frequency restoration. Compared to baselines, our VL-DUN effectively distinguishes between noise and texture, preserving more details.}
  \label{fig:fre}
\end{figure}

The quality result is shown in Table~\ref{tab:quant_comparison}, demonstrating our VL-DUN achieves state-of-the art by improving 0.92dB in PSNR and 9.76\% in Dice. The visual results are provided in Fig.~\ref{fig:visual} and Fig.~\ref{fig:fre}, showing that our VL-DUN reconstructs in more details. This substantial improvement validates that our joint optimization strategy effectively leverages the synergy between tasks: the restoration branch provides clearer features for segmentation, while the segmentation branch imposes semantic constraints to prevent structural distortion during restoration. To provide a deeper understanding of this capability, we visualize the high-frequency components in Fig.~\ref{fig:fre}. It is evident that our model, empowered by the Frequency-Aware Mamba mechanism, successfully decouples noise from texture. This allows VL-DUN to suppress artifacts while preserving high-frequency diagnostic details, addressing the spectral bias limitation common in conventional Mamba-based architectures.

\subsection{AiOMedIS task}

We carry out comparative experiments on AiOMedIS. Baseline models are SAM~\cite{kirillov2023segment}, MedSAM~\cite{ma2024segment}, EM-DenoiSeg~\cite{DBLP:conf/miccai/WangLCSDHX24}, IMIS~\cite{cheng2024interactivemedicalimagesegmentation}, SAM3~\cite{carion2025sam3segmentconcepts}.

\begin{figure*}[t]
  \centering
  \includegraphics[width=1\textwidth]{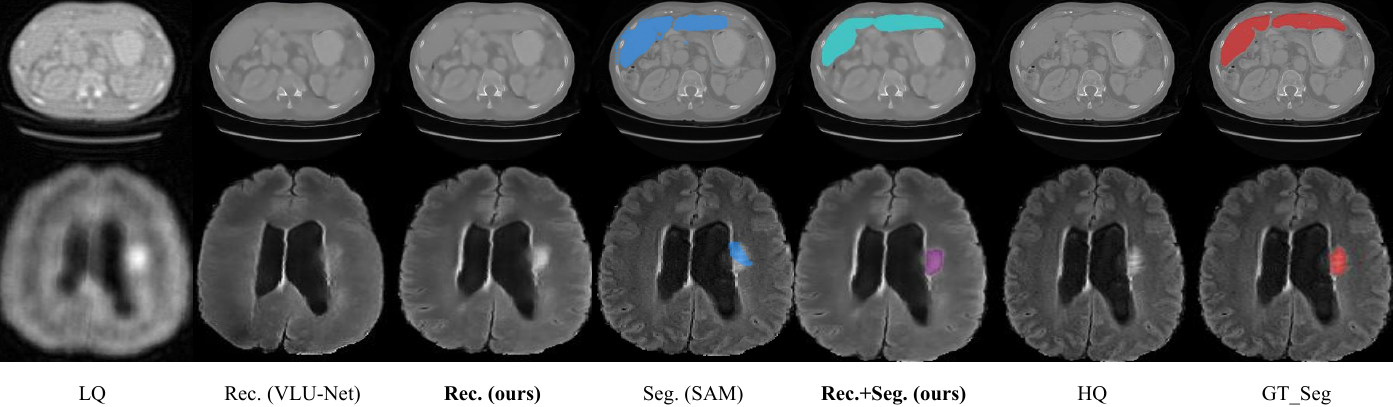} 
  \caption{Visual results of our VL-DUN complete joint restoration and segmentation task. Our VL-DUN simultaneously completes reconstruction and segmentation tasks, performing excellently on multimodal datasets. We presented the reconstruction results of VLUNet and the segmentation results of SAM for comparison, further demonstrating the superiority of our model.}
  \label{fig:seg}
\end{figure*}

Table \ref{tab:seg_comparison} reveals that our method outperforms specialist segmentation models such as SAM and Med-SAM. It should be emphasized that the input to the specially designed segmentation model is always HQ images, whereas the input to VL-DUN is LQ images, and its output includes the segmented mask and the reconstructed image. As shown in Fig.~\ref{fig:seg}, our VL-DUN successfully accomplishes the joint tasks of restoration and segmentation, providing a feasible paradigm for the AiOMIRS task.

\begin{table}[t]
    \centering
    \caption{\textbf{Segmentation performance comparison on four datasets.} We compare the Dice coefficient and IoU with state-of-the-art segmentation methods. \best{Red} and \second{Blue} indicate the best and second-best performance, respectively.}
    \label{tab:seg_comparison}
    \resizebox{0.5\textwidth}{!}{
        \begin{tabular}{c|cc|cc|cc|cc|cc}
            \toprule
            \multirow{2}{*}{\textbf{Method}} & \multicolumn{2}{c|}{\textbf{ACDC}} & \multicolumn{2}{c|}{\textbf{COVID19CTscans}} & \multicolumn{2}{c|}{\textbf{HCC-TACE-Seg}} & \multicolumn{2}{c|}{\textbf{BraTS2021}} & \multicolumn{2}{c}{\textbf{Average}} \\
            \cmidrule(lr){2-3} \cmidrule(lr){4-5} \cmidrule(lr){6-7} \cmidrule(lr){8-9} \cmidrule(lr){10-11}
             & Dice$\uparrow$ & IoU$\uparrow$ & Dice$\uparrow$ & IoU$\uparrow$ & Dice$\uparrow$ & IoU$\uparrow$ & Dice$\uparrow$ & IoU$\uparrow$ & Dice$\uparrow$ & IoU$\uparrow$ \\
            \midrule
            SAM & {0.6891} & \second{0.6408} & 0.6588 & 0.6170 & 0.5343 & 0.4187 & \best{0.5343} & \best{0.4810} & 0.6041 & 0.5394 \\
            
            Med-SAM & 0.6754 & 0.5635 & 0.6418 & 0.5475 & 0.5564 & 0.4574 & \second{0.5043} & 0.3998 & 0.5945 & 0.4921 \\
            
            EM-DenoiSeg & 0.6215 & 0.6321 & 0.5947 & 0.5123 & 0.5951 & 0.5647 & 0.3945 & 0.3945 & 0.5515 & 0.5259 \\
            
            IMIS & \second{0.7283} & 0.6189 & 0.5605 & 0.4328 & \best{0.7577} & \best{0.6749} & 0.5035 & 0.4147 & \second{0.6375} & 0.5353 \\
            
            SAM3 & 0.6825 & 0.5682 & \second{0.6874} & \second{0.6177} & \second{0.6721} & 0.6024 & 0.5015 & \second{0.4287} & 0.6359 & \second{0.5543} \\
            
            \rowcolor{gray!20} \textbf{VL-DUN (Ours)} & \best{0.7563} & \best{0.6630} & \best{0.7742} & \best{0.7410} & 0.6401 & \second{0.6054} & 0.4680 & 0.4047 & \best{0.6597} & \best{0.6035} \\
            \bottomrule
        \end{tabular}
    }
\end{table}

\begin{table*}[t]
    \centering
    \caption{\textbf{Ablation study of different components in VL-DUN.} We investigate the contribution of the CLIP embedding, CLIP fine-tuning, Frequency Decoupling (FD), the Segmentation Head, and the Reconstruction Head. \best{Red} and \second{Blue} indicate the best and second-best performance, respectively. Ablation experiments demonstrated that segmentation and reconstruction tasks can mutually enhance each other, and that all of our components and strategies are effective.}
    \label{tab:ablation_study}
    \resizebox{\textwidth}{!}{
        \begin{tabular}{c|ccc|ccc|ccc|ccc|ccc}
            \toprule
            \multirow{2}{*}{\textbf{Method}} & \multicolumn{3}{c|}{\textbf{ACDC}} & \multicolumn{3}{c|}{\textbf{COVID19CTscans}} & \multicolumn{3}{c|}{\textbf{HCC-TACE-Seg}} & \multicolumn{3}{c|}{\textbf{BraTS2021}} & \multicolumn{3}{c}{\textbf{Average}} \\
            \cmidrule(lr){2-4} \cmidrule(lr){5-7} \cmidrule(lr){8-10} \cmidrule(lr){11-13} \cmidrule(lr){14-16}
             & PSNR$\uparrow$ & SSIM$\uparrow$ & Dice$\uparrow$ & PSNR$\uparrow$ & SSIM$\uparrow$ & Dice$\uparrow$ & PSNR$\uparrow$ & SSIM$\uparrow$ & Dice$\uparrow$ & PSNR$\uparrow$ & SSIM$\uparrow$ & Dice$\uparrow$ & PSNR$\uparrow$ & SSIM$\uparrow$ & Dice$\uparrow$ \\
            \midrule
            \rowcolor{gray!20} \textbf{VL-DUN (Ours)} & \best{25.93} & \best{0.7129} & \best{0.7563} & \best{22.40} & \best{0.5330} & \best{0.7742} & \best{35.63} & 0.9239 & \best{0.6401} & \second{32.31} & \second{0.8625} & \best{0.4680} & \best{29.07} & \second{0.7581} & \best{0.6597} \\
            
            w/o CLIP & 24.42 & 0.6794 & 0.7032 & 21.98 & 0.5208 & 0.6143 & 30.41 & 0.8945 & 0.5124 & 32.01 & 0.8514 & 0.4068 & 27.21 & 0.7365 & 0.5592 \\
            
            w/o CLIP tuning & 25.32 & 0.7010 & \second{0.7045} & 20.98 & 0.5295 & \second{0.7512} & \second{35.21} & \best{0.9271} & 0.6031 & \best{32.65} & \best{0.8783} & \second{0.4453} & \second{28.54} & \best{0.7590} & \second{0.6260} \\
            
            w/o FD & 24.45 & 0.6812 & 0.6842 & 21.62 & 0.5182 & 0.6451 & 33.41 & 0.8984 & 0.6143 & 32.21 & 0.8551 & 0.3951 & 27.92 & 0.7382 & 0.5847 \\
            
            w/o Seg Head & \second{25.52} & \second{0.7079} & - & \second{22.32} & 0.5277 & - & 33.56 & \second{0.9255} & - & 32.21 & 0.8571 & - & 28.40 & 0.7546 & - \\
            
            w/o Rec Head & - & - & 0.7031 & - & - & 0.7315 & - & - & \second{0.6214} & - & - & 0.4231 & - & - & 0.6198 \\
            \bottomrule
        \end{tabular}
    }
\end{table*}

\subsection{Ablation Study}

In this section, we conduct a comprehensive ablation study to validate the effectiveness of each component in VL-DUN, analyzing the impact of vision-language priors, frequency decoupling, and the mutual synergy between restoration and segmentation tasks.

As detailed in Table \ref{tab:ablation_study}, the removal of the CLIP module (\textit{w/o CLIP}) results in the most significant performance drop, confirming that semantic guidance is critical for preventing blind restoration. Furthermore, the \textit{w/o FD} variant exhibits a noticeable decline in both SSIM and Dice scores. This supports our hypothesis regarding Mamba's spectral bias, proving that our Frequency-Aware design effectively preserves fine-grained pathological details by decoupling high-frequency texture processing.

We investigate the mutual reinforcement between tasks by removing the segmentation head and the reconstruction head respectively. \textbf{Restoration benefits from Segmentation:} Contrary to the common trade-off where multi-task learning degrades individual task performance, our full model achieves higher restoration quality compared to the reconstruction-only variant. This suggests that the segmentation task acts as a semantic regularizer, forcing the encoder to learn structure-aware features that benefit image restoration. \textbf{Segmentation benefits from Restoration:} When the restoration branch is removed, the segmentation performance drops significantly. This demonstrates that the restoration branch provides essential supervision for feature denoising. Without this explicit restoration objective, the shared encoder struggles to extract clean features from degraded inputs, leading to inferior segmentation results. In addition, we further discuss the principles and experimental demonstrations of the synergy between reconstruction and segmentation tasks in Appendix~\ref{sec:synergy_analysis}.

\section{Conclusion}
In this paper, we present VL-DUN, a unified framework that redefines the relationship between medical image restoration and downstream analysis and provides a paradigm for AiOMIRS task. 
We introduce a Vision-Language Dual Prior mechanism. By aligning low-level optimization with high-level semantic descriptions, we enable the model to cognitively adapt to diverse imaging modalities and degradation patterns, effectively narrowing the gap between varying clinical protocols.
Second, we tackle the inherent spectral bias of Mamba. Our proposed Frequency-Aware Mamba strategy successfully reconciles the conflict between long-range dependency modeling and fine-grained texture preservation, offering a solution that does not compromise on high-frequency details.
Experimental results are conclusive: VL-DUN outperforms state-of-the-art baselines with significant margins (0.92 dB in PSNR and 9.76\% in Dice). Beyond metrics, we prove that restoration and segmentation are mutually beneficial, unlocking a new paradigm where high-level semantics and low-level vision synergistically reinforce each other for robust clinical analysis.

\section*{Acknowledgements}

Our work is based on IMIS-bench, AMIR, and VLUNet. We are grateful for their outstanding work and open-source contributions.

\nocite{langley00}

\bibliography{example_paper}
\bibliographystyle{icml2026}

\newpage
\appendix
\onecolumn

\section{Challenges in AiOMIRS Task}

The All-in-One Medical Image Restoration and Segmentation (AiOMIRS) task is not merely a summation of two independent problems. It presents a unique set of challenges stemming from the heterogeneity of medical data, the ill-posed nature of blind restoration, and the inherent conflict between low-level and high-level optimization objectives. In this section, we discuss these core challenges to highlight the motivation of our VL-DUN.

\subsection{The data distribution between each modality is inconsistent.}

The non-i.i.d. nature of multi-modal medical data (e.g., CT vs. MRI) creates a disjoint feature space\cite{yang2024all}, making it difficult for a static network to capture shared representations without suffering from domain shift. This motivates our \textbf{Vision-Language Dual Prior}, which explicitly injects modality information to condition the network, effectively disentangling the heterogeneous distributions.

\subsection{Gradient Interference and Negative Transfer}

As illustrated by pioneer AiOMedIR work AMIR~\cite{yang2024all}, during joint training, the optimization direction for one modality (or task) often conflicts with another. This \textit{gradient interference} leads to negative transfer, where the performance on specific modalities degrades as the model capacity is split to accommodate conflicting patterns. VL-DUN addresses this by using CLIP-based text priors to perform \textbf{dynamic operator estimation} (via the AMGDM module), allowing the network to "perceive" the degradation type before restoring it.

\subsection{Objective Misalignment between Restoration and Segmentation}

Fundamentally, restoration and segmentation impose contradictory spectral demands on feature representation. Restoration tasks necessitate the preservation of high-frequency components to ensure textural fidelity and edge sharpness. In contrast, semantic segmentation typically operates as a low-pass filter\cite{DBLP:conf/icml/RahamanBADLHBC19}, abstracting away local details to capture contiguous anatomical shapes. A naive unified model, which enforces a shared feature space for both tasks, struggles to reconcile these opposing \textit{inductive biases}. Without explicit decoupling mechanisms, the network tends to converge to a suboptimal compromise—either over-smoothing textures to satisfy segmentation or retaining excessive noise that disrupts semantic consistency.

To resolve this feature-level conflict, our VL-DUN introduces a \textbf{Frequency-Aware Mamba} mechanism within the unfolding stages. By internally decoupling feature processing, we explicitly assign the recovery of high-frequency textures to the Bi-Mamba branch and the preservation of low-frequency structural information to the GDFN branch. This design ensures that the learned representations are sufficiently rich for high-fidelity restoration yet structurally coherent for precise segmentation, effectively reconciling the spectral discrepancy within a unified framework.

\section{Theoretical and Practical Analysis of Task Synergy}
\label{sec:synergy_analysis}

A core motivation of VL-DUN is the hypothesis that low-level restoration and high-level semantic segmentation are mutually beneficial. Here, we provide a rigorous analysis grounded in Bayesian inference and solution space geometry to explain this closed-loop synergy.

\subsection{Restoration Promotes Segmentation}
In the All-in-One setting, the input $y$ is corrupted by heterogeneous degradations, modeled as $y = \Phi x + \mathbf{n}$. Direct segmentation attempts to model the posterior probability $P(S|y)$, where $S$ is the semantic mask. However, due to the data processing inequality, the information content about the semantic structure $S$ is degraded by noise and the operator $\Phi$, \textit{i.e.}, $I(S; y) \leq I(S; x)$, where $I(\cdot;\cdot)$ denotes Mutual Information.

By integrating the restoration module (DUN) as a pre-processor, we aim to recover an estimate $\hat{x}$ that maximizes the log-likelihood of the clean signal distribution $\log P(x|y)$. This process mathematically transforms the conditional distribution:
\begin{equation}
    P(S|y) \approx \int P(S|\hat{x}) P(\hat{x}|y) d\hat{x}.
\end{equation}
In a direct segmentation model, the term $P(\hat{x}|y)$ is implicitly modeled with high variance due to the ill-posed nature of degradations. In our VL-DUN, the restoration module explicitly minimizes the reconstruction error $\|\hat{x} - x\|^2$, effectively collapsing the distribution $P(\hat{x}|y)$ towards a Dirac delta function centered at the true manifold $\mathcal{M}_{clean}$. This domain alignment minimizes the \textit{aleatoric uncertainty} $\mathcal{H}(S|y)$ caused by noise, allowing the segmentation head to approximate the ideal posterior $P(S|x)$, thereby tracing true anatomical boundaries rather than artifact-induced noise.

\subsection{Segmentation Promotes Restoration: Reducing the Solution Space via Semantic Regularization}
The synergy from high-level semantics to low-level restoration can be formulated as a constraint satisfaction problem that reduces the ill-posedness of the inverse problem.

The restoration task aims to find a solution $\hat{x}$ within the feasible set defined by data fidelity:
\begin{equation}
    \Omega_{data} = \{x \mid \|y - \Phi x\|^2 \leq \epsilon \}.
\end{equation}
Since $\Phi$ is often rank-deficient (e.g., in undersampling or blur), the null space $\mathcal{N}(\Phi)$ is non-empty. Any component $x_{null} \in \mathcal{N}(\Phi)$ can be added to the solution without affecting data fidelity, making $\Omega_{data}$ a vast solution space containing infinite artifacts (the "ill-posed" nature). Traditional regularizers like $\mathcal{L}_{1}$ or TV constrain this space to a smooth manifold $\Omega_{smooth}$, which often leads to the over-smoothing of weak textures.

In contrast, the segmentation task introduces a Semantic Manifold Constraint. Let $\mathcal{L}_{seg}$ be the segmentation loss. We define the semantic solution space as:
\begin{equation}
    \Omega_{sem} = \{x \mid \nabla \mathcal{L}_{seg}(x) \text{ aligns with anatomical boundaries} \}.
\end{equation}
Our joint optimization seeks the intersection $\hat{x} \in \Omega_{data} \cap \Omega_{sem}$.
Since $\Omega_{sem}$ enforces piecewise consistency and sharp transitions at object borders, the intersection space is significantly smaller than $\Omega_{data}$ alone: $|\Omega_{data} \cap \Omega_{sem}| \ll |\Omega_{data}|$.
This effectively "locks" the high-frequency components corresponding to semantic edges, preventing them from falling into the null space of the degradation operator.

\begin{figure*}[htbp]
  \centering
  \includegraphics[width=1\textwidth]{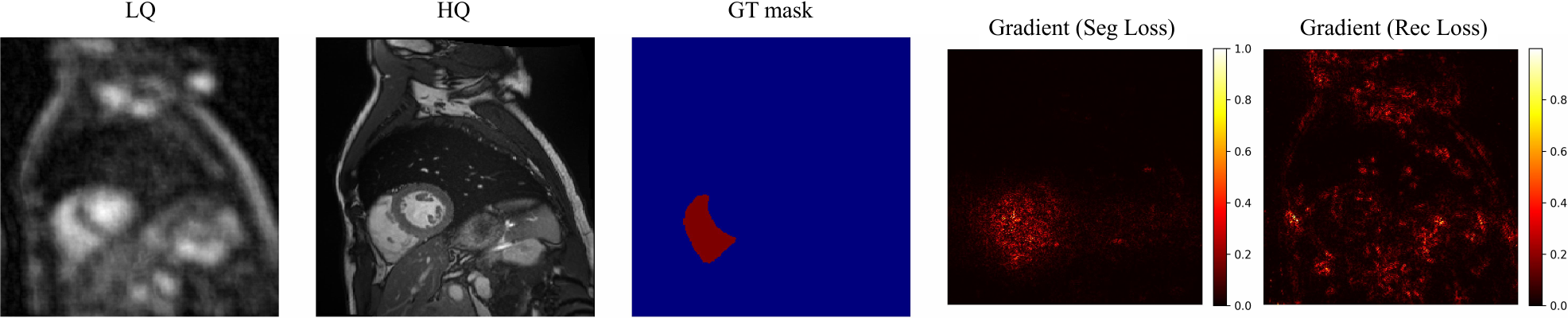} 
  \caption{Visual evidence of task interaction via gradient saliency maps.}
  \label{fig:gradient_vis}
\end{figure*}

Mathematically, considering the gradient descent update, the segmentation loss acts as a dynamic structural prior:
\begin{equation}
    x_{t+1} \leftarrow x_t - \eta (\underbrace{\nabla \mathcal{L}_{rec}}_{\text{Texture Fidelity}} + \lambda \underbrace{\nabla \mathcal{L}_{seg}}_{\text{Shape Constraint}}).
\end{equation}
As visualized in Figure~\ref{fig:gradient_vis}, $\nabla \mathcal{L}_{seg}$ exhibits a sparse, edge-aware pattern:
\begin{itemize}
    \item \textbf{Texture Suppression (Null-Space Filtering):} Inside homogeneous organs, $\nabla \mathcal{L}_{seg} \approx 0$. This implies the segmentation loss does not interfere with the restoration of intra-region textures, avoiding the introduction of false hallucinations.
    \item \textbf{Boundary Sharpening (Edge Constraint):} At anatomical borders, $\|\nabla \mathcal{L}_{seg}\|$ is maximized. This creates a high-energy barrier against smoothing operations, forcing the Deep Unfolding Network to preserve crisp edges even if the data fidelity term (e.g., MSE) encourages blurring.
\end{itemize}
Thus, the segmentation head functions not just as a multi-task objective, but as a \textbf{semantic regularizer that reduces the solution space} to physically plausible anatomical structures.

\section{Why use VLM (CLIP) to extract semantic information instead of other methods}
\label{sec:CLIP}
In the context of AiOMIRS, accurately identifying the modality and degradation type is a prerequisite for effective restoration. We explain the choice of CLIP over traditional CNN classifiers or VAE-based latent representations.

\subsection{Limitations in Traditional CNNs and VAEs}

Latest works such as RealDGen~\cite{peng2025towards,CheHao_AllinOne_MICCAI2025}, use VAEs and CNNs to extract content or degradation features of different images. Although they adopt contrastive learning methods or other effective approaches to enable models to extract semantic information from images, the following issues arise in the context of medical imaging:

\begin{figure*}[htbp]
  \centering
  \includegraphics[width=1\textwidth]{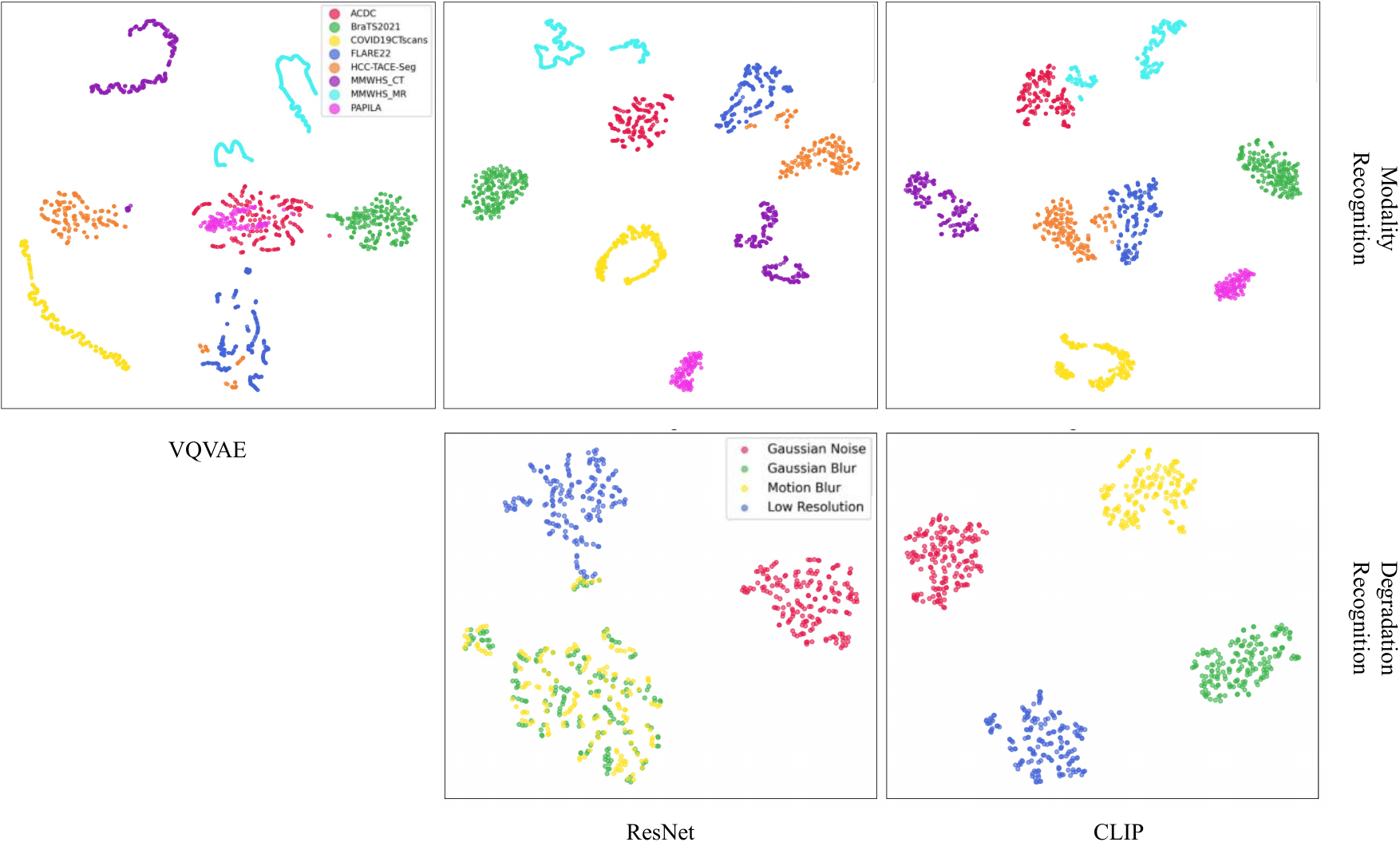} 
  \caption{We compare the CLIP with VQVAE and ResNet. CLIP has strong ability to extract modality and degradation features.}
  \label{fig:extract}
\end{figure*}

\begin{itemize}
    \item \textbf{Prior information is biased.} CNNs require large amounts of data to learn image features, especially in scenarios involving different modalities. However, due to the imbalance in medical datasets, the features learned by CNNs can be biased. As for VAEs, most applications are in natural images, and large-scale fine-tuning has not been performed on multi-modal medical images, which can also lead to feature bias. 
    \item \textbf{Features are compressed.} Traditional VAEs typically compress the features of an image into a latent space. Although most of the effective features of the image are preserved, in all-in-one medical imaging tasks, even if the images are all MRI or CT scans, they may become different modalities due to different imaging mechanisms. In such cases, traditional VAEs require a large number of dimensions to distinguish these features, which increases the model size and training difficulty.
    \item \textbf{Weak semantic information.} Unlike CLIP, CNNs and VAEs are purely visual approaches that process images through a series of operations and distinguish different features using methods like contrastive learning. However, semantic features obtained from this single-modality approach have weak interpretability and poor generalization.
\end{itemize}

\subsection{The superiority of CLIP}

In contrast, we leverage CLIP to extract semantic features for the following reasons:

\begin{itemize}
    \item \textbf{Rich prior knowledge.} CLIP is pre-trained extensively on a large number of images and can already recognize modal information and degradation features fairly well without fine-tuning, as shown in Fig.~\ref{fig:Modality_Finetune} and Fig.~\ref{fig:Degradation_Finetune}. This results in low fine-tuning costs for us. At the same time, with the current maturity of LoRA fine-tuning methods~\cite{zanella2024low}, the cost of fine-tuning CLIP is significantly better than training VAEs and CNNs from scratch.
    \item \textbf{Language features as a reference for alignment.} Although the features of natural images and medical images differ significantly, the degraded information is consistent in the language dimension, which provides a strong reference when we perform contrastive learning.
\end{itemize}

Further, we conduct experiments to validate our analysis. As shown in Fig.~\ref{fig:extract}, we use CLIP to extract modality features comparing to ResNet~\cite{DBLP:conf/cvpr/HeZRS16} and VQVAE~\cite{DBLP:conf/nips/OordVK17}. We also use CLIP to extract degradation features compared to ResNet~\cite{DBLP:conf/cvpr/HeZRS16}. To ensure a fair comparison, we fine-tuned all three models for the same epochs.

\section{Justification for Degradation Prior Selection}
\label{sec:degradation_justification}

A critical design choice in VL-DUN is fine-tuning the CLIP degradation prior on fundamental degradation types (e.g., noise, blur, low-resolution) rather than explicit physical models like k-space undersampling. This decision is grounded in the hypothesis of \textit{Visual Primitives} and the flexibility of our attention mechanism. Here, we provide a detailed justification from three perspectives.

\subsection{Decomposition into Visual Primitives}
Complex physical degradations often manifest visually as combinations of fundamental artifacts. We posit that fundamental degradations serve as orthogonal \textbf{visual primitives} (or basis vectors) that span the space of medical imaging artifacts.
\begin{itemize}
    \item \textbf{CT Streak Artifacts:} In sparse-view CT reconstruction, the resulting streak artifacts structurally resemble high-frequency, directional noise patterns. In the semantic embedding space of CLIP, these share significant overlap with ``Structured Noise'' or ``Motion Blur''.
    \item \textbf{MRI Aliasing:} Undersampling in MRI k-space results in aliasing ghosts and blurring. Visually, these can be decomposed into a mixture of ``Defocus Blur'' (loss of high frequencies) and ``Noise'' (incoherent aliasing artifacts).
\end{itemize}
By training on these primitives, the network learns to represent complex, domain-specific artifacts (like undersampling) as a weighted combination of these fundamental embeddings, rather than treating them as entirely new, unseen categories.

\subsection{Soft-Mapping via Cross-Attention}
Crucially, our \textbf{Attention-Mamba Gradient Descent Module (AMGDM)} does not perform hard classification (i.e., it does not force the model to categorize an image strictly as ``Blur'' or ``Noise''). Instead, it utilizes a \textbf{Cross-Attention mechanism}:
\begin{equation}
    \text{Attention}(Q, K, V) = \text{softmax}\left(\frac{Q K^\top}{\sqrt{d_k}}\right) V
\end{equation}
where the degraded image features ($Q$) query the CLIP priors ($K, V$). Even without an explicit ``Undersampling'' prompt, the attention mechanism dynamically computes similarity scores based on feature affinity. For instance, as visualized in our attention maps, the model successfully attends to CT streaks by leveraging the ``Noise'' and ``Blur'' priors as \textit{feature anchors}, effectively using them to localize high-frequency anomalies regardless of the semantic label.

\subsection{Generalization across Modalities}
Physical undersampling models vary drastically between modalities. For example, artifacts arise from the Radon transform in CT versus Fourier transform aliasing in MRI. Training CLIP on specific physical models would likely cause the prior to overfit to a specific imaging physics, limiting its transferability.
Using universal visual degradations allows us to decouple the semantic prior from specific imaging physics. This ensures that our VL-DUN remains \textbf{modality-agnostic}, capable of handling heterogeneous medical tasks within a unified framework, which aligns with the core philosophy of the AiOMIRS task.

\section{Detailed Components of our DUN}
\label{sec:dun_ab}

\begin{figure*}[htbp]
  \centering
  \includegraphics[width=1\textwidth]{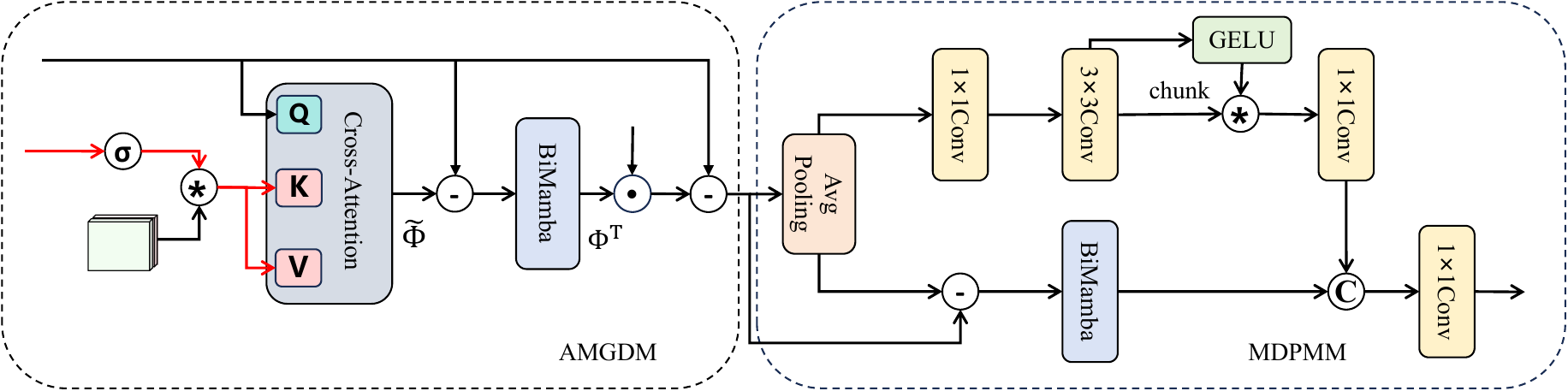} 
  \caption{Detailed components of our AMGDM and MDPMM.}
  \label{fig:pmm}
\end{figure*}

Our VL-DUN is constructed by unfolding the iterative Proximal Gradient Descent (PGD) algorithm into a deep neural network with $K$ cascaded stages. Each stage $k$ consists of two synergistic modules: the \textbf{Attention-Mamba Gradient Descent Module (AMGDM)} for data fidelity update, and the \textbf{Mamba-GDFN based Proximal Map Module (MDPMM)} for denoising.

\subsection{Attention-Mamba Gradient Descent Module (AMGDM)}
\begin{figure*}[htbp]
  \centering
  \includegraphics[width=1\textwidth]{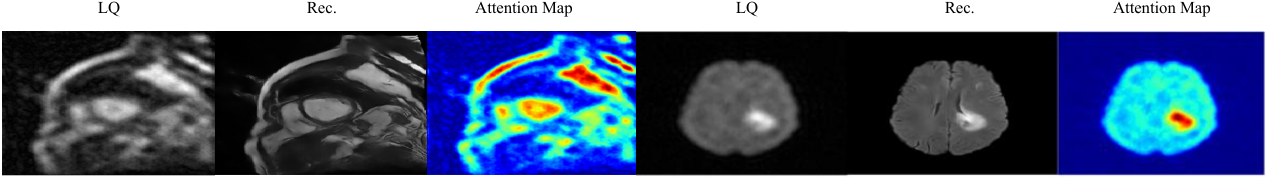} 
  \caption{Visualization of the Attention Map.}
  \label{fig:attention}
\end{figure*}

The AMGDM is designed to solve the data fidelity term $\frac{1}{2}\|y - \Phi x\|^2$ by dynamically estimating the degradation operator $\Phi$.
As illustrated in Figure~\ref{fig:pmm}, we introduce a Vision-Language Prior Injection mechanism:
\begin{itemize}
    \item \textbf{Operator Estimation via Cross-Attention.} We treat the image features as Queries ($Q$) and the CLIP-extracted dual priors (Modality \& Degradation) as Keys ($K$) and Values ($V$). This Cross-Attention mechanism dynamically calibrates the degradation operator $\Phi$ based on the specific artifact type described by the priors. We provide visualization of the Attention Map in Fig.~\ref{fig:attention}, which demonstrates the powerful ability of Cross-Attention.
    \item \textbf{Gradient Back-projection via Bi-Mamba.} To approximate the transpose operator $\Phi^\top$ (which maps the residual error back to the image space), we utilize a Bi-directional Mamba layer. This allows for global error accumulation with linear computational complexity $\mathcal{O}(N)$, ensuring efficient gradient updates.
\end{itemize}

\subsection{Frequency-Aware Proximal Module (MDPMM)}
The MDPMM acts as the proximal operator to project the intermediate estimate onto the clean image manifold. Addressing the spectral bias of standard Mamba models (which tend to filter out high-frequency details), we propose a Frequency Decoupling Strategy.
\begin{itemize}
    \item \textbf{Frequency Decomposition.} The input feature is split into low-frequency ($x_{low}$) and high-frequency ($x_{high}$) components via average pooling and subtraction.
    \item \textbf{Dual-Branch Processing.} 
    (1) The \textbf{High-Frequency Branch} employs a Bi-Mamba layer to capture global dependencies within the texture components, distinguishing noise from fine anatomical details.
    (2) The \textbf{Low-Frequency Branch} utilizes a Gated-Dconv Feed-Forward Network (GDFN) to preserve stable structural information.
\end{itemize}
The outputs are fused to update the latent image $x_k$, ensuring both structural consistency and texture fidelity. It must be mentioned that our Seg. Head is made up of a simple CNN.

\begin{table*}[t]
    \centering
    \caption{\textbf{Segmentation performance comparison on four datasets.} We compare the Dice coefficient and IoU with state-of-the-art segmentation methods. \best{Red} and \second{Blue} indicate the best and second-best performance, respectively.}
    \label{tab:niu}
    \resizebox{1\textwidth}{!}{
        \begin{tabular}{c|cc|cc|cc|cc|cc}
            \toprule
            \multirow{2}{*}{\textbf{Method}} & \multicolumn{2}{c|}{\textbf{ACDC}} & \multicolumn{2}{c|}{\textbf{COVID19CTscans}} & \multicolumn{2}{c|}{\textbf{HCC-TACE-Seg}} & \multicolumn{2}{c|}{\textbf{BraTS2021}} & \multicolumn{2}{c}{\textbf{Average}} \\
            \cmidrule(lr){2-3} \cmidrule(lr){4-5} \cmidrule(lr){6-7} \cmidrule(lr){8-9} \cmidrule(lr){10-11}
             & Dice$\uparrow$ & IoU$\uparrow$ & Dice$\uparrow$ & IoU$\uparrow$ & Dice$\uparrow$ & IoU$\uparrow$ & Dice$\uparrow$ & IoU$\uparrow$ & Dice$\uparrow$ & IoU$\uparrow$ \\
            \midrule
            SAM & {0.6891} & \second{0.6408} & 0.6588 & 0.6170 & 0.5343 & 0.4187 & \best{0.5343} & \best{0.4810} & 0.6041 & 0.5394 \\
            
            Med-SAM & 0.6754 & 0.5635 & 0.6418 & 0.5475 & 0.5564 & 0.4574 & \second{0.5043} & 0.3998 & 0.5945 & 0.4921 \\
            
            EM-DenoiSeg & 0.6215 & 0.6321 & 0.5947 & 0.5123 & 0.5951 & 0.5647 & 0.3945 & 0.3945 & 0.5515 & 0.5259 \\
            
            IMIS & \second{0.7283} & 0.6189 & 0.5605 & 0.4328 & \best{0.7577} & \best{0.6749} & 0.5035 & 0.4147 & \second{0.6375} & 0.5353 \\
            
            SAM3 & 0.6825 & 0.5682 & \second{0.6874} & \second{0.6177} & \second{0.6721} & 0.6024 & 0.5015 & \second{0.4287} & 0.6359 & \second{0.5543} \\
            
            \rowcolor{gray!20} \textbf{VL-DUN (Ours)} & \best{0.7563} & \best{0.6630} & \best{0.7742} & \best{0.7410} & 0.6401 & \second{0.6054} & 0.4680 & 0.4047 & \best{0.6597} & \best{0.6035} \\
            \bottomrule
        \end{tabular}
    }
\end{table*}

\section{Visual results of the intermediate features provided by our VL-DUN}

\begin{figure*}[htbp]
  \centering
  \includegraphics[width=1\textwidth]{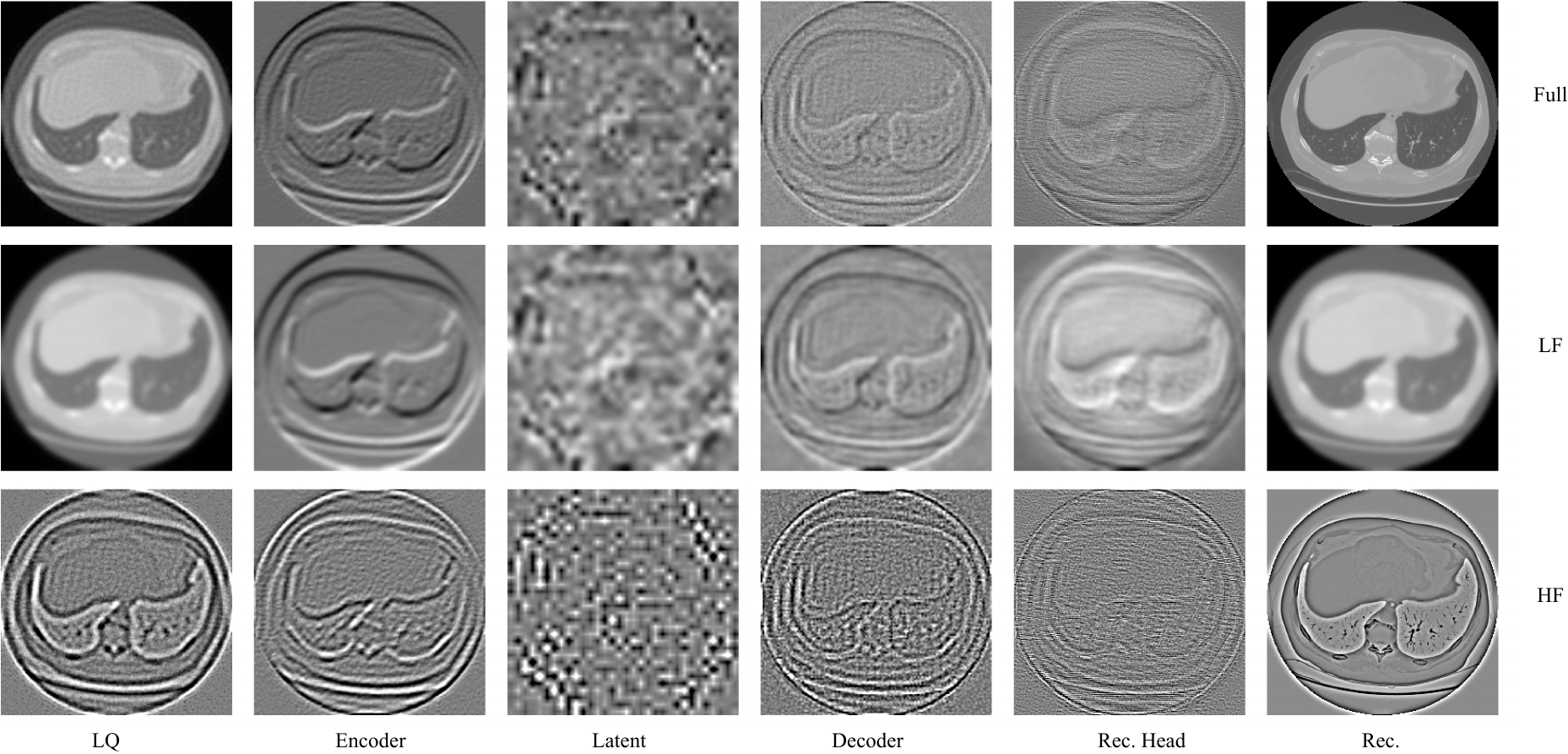} 
  \caption{We thoroughly visualize the features of each layer of the network, showing the complete reconstruction process, the reconstruction process of low-frequency features, and the reconstruction process of high-frequency features.}
  \label{fig:interpret}
\end{figure*}

\noindent\textbf{Interpretability and Frequency Analysis.} 
Unlike conventional deep learning models, our VL-DUN offers a transparent view of the restoration process, inherently rooted in the iterative Deep Unfolding framework. As illustrated in Figure~\ref{fig:interpret}, we visualize the intermediate characteristic evolution across three frequency bands:
\begin{itemize}
    \item \textbf{Full Spectrum (Top Row).} Demonstrates the gradual refinement from the Low-Quality (LQ) input to the High-Quality (HQ) output, verifying the effectiveness of the stage-wise optimization.
    \item \textbf{Low-Frequency Branch (Middle Row).} Focuses on recovering global anatomical structures and contrast distribution. The visualization confirms that the GDFN module stably preserves organ shapes and smooth regions throughout the encoding and decoding stages.
    \item \textbf{High-Frequency Branch (Bottom Row).} Highlights the model's ability to distinguish noise from fine textures. The Bi-Mamba module effectively suppresses artifacts in the latent space while progressively enhancing edge details in the reconstruction head.
\end{itemize}
This visualization explicitly validates our Frequency-Aware design, proving that the network learns to decouple structural preservation from texture restoration, providing clinicians with a interpretable reconstruction trajectory.

\section{Results of AiOMIRS}

The larger version of Tab~\ref{tab:seg_comparison} is shown in Table~\ref{tab:niu}.

\section{Computational Reports}

Our VL-DUN has the lowest number of parameters among all-in-one models, even fewer than AMIR. At the same time, its FLOPS are the lowest among all all-in-one models, and the inference time is second only to PromptIR among all-in-one models, further demonstrating the superiority of our model. 

\begin{table}[htbp]
    \centering
    \caption{\textbf{Complexity analysis.} We compare the number of parameters, FLOPs, and inference time with state-of-the-art methods. FLOPs are calculated on an input size of $256 \times 256$. The gray row highlights our proposed method.}
    \label{tab:complexity}
    \resizebox{0.5\textwidth}{!}{
        \begin{tabular}{c|ccc}
            \toprule
            \textbf{Method} & \textbf{Param. (M)} & \textbf{FLOPs (G)} & \textbf{Time (ms)} \\
            \midrule
            DenoiSeg & 1.93 & 20.95 & 3.38 \\
            PromptIR & 34.11 & 281.64 & 52.44 \\
            AMIR & 23.54 & 254.11 & 76.96 \\
            EM-DenoiSeg & 5.51 & 71.46 & 7.74 \\
            VLU-Net & 35.35 & 335.80 & 82.67 \\
            TAT & 32.19 & 259.88 & 175.81 \\
            \rowcolor{gray!20} \textbf{VL-DUN (Ours)} & 22.77 & 167.24 & 58.72 \\
            \bottomrule
        \end{tabular}
    }
\end{table}

\section{Limitations and Future Works}

As a pioneering work in AiOMIRS, we have established a new benchmark and explored the synergy between restoration and segmentation. While VL-DUN demonstrates superior performance, we acknowledge several limitations that point toward future research directions:

\begin{itemize}
    \item \textbf{Limited Physical Priors in VLM.} While CLIP possesses powerful semantic extraction capabilities suitable for degradation recognition, it lacks explicit knowledge of medical imaging physics (e.g., MRI k-space trajectories or CT beam hardening). Its representations are inherently semantic rather than physically interpretable. Future work could explore aligning VLMs with physics-informed encoders to bridge this gap.
    \item \textbf{Gap in Degradation Simulation.} Our training relies on physically grounded simulation to generate LQ data from HQ references. Although we adhere to imaging principles (e.g., Fourier undersampling), there remains a domain gap compared to real-world clinical artifacts where paired data is unavailable.
    \item \textbf{Simplified Segmentation Head.} To prioritize computational efficiency and maintain a lightweight architecture, we employed a basic convolutional network as the segmentation head. While effective, this design may not fully handle the extreme class imbalance in certain medical datasets. Future iterations will investigate more dedicated, yet efficient, segmentation heads to address this challenge.
\end{itemize}


\end{document}